\documentclass{article}
\usepackage{graphicx}
\usepackage{comment}
\usepackage{color}
\usepackage{url}
\usepackage[utf8]{inputenc}

\usepackage{authblk}

\begin{document}
\title{The Block-based Mobile PDE Systems Are Not Secure – Experimental Attacks}
\date{}
\author[1]{\small Niusen Chen}
\author[1]{\small Bo Chen}
\author[2]{\small Weisong Shi}
\affil[1]{Department of Computer Science, Michigan Technological University, Michigan, United States}
\affil[2]{Department of Computer Science, Wayne State University, Michigan, United States}
\renewcommand*{\Affilfont}{\small}

\maketitle
\begin{abstract}
Nowadays, mobile devices have been used broadly to store and process sensitive data. To ensure confidentiality of the sensitive data, Full Disk Encryption (FDE) is often integrated in mainstream mobile operating systems like Android and iOS. FDE however cannot defend against coercive attacks in which the adversary can force the device owner to disclose the decryption key. To combat the coercive attacks, Plausibly Deniable Encryption (PDE) is leveraged to plausibly deny the very existence of sensitive data. However, most of the existing PDE systems for mobile devices are deployed at the block layer and suffer from deniability compromises. 

Having observed that none of existing works in the literature have experimentally demonstrated the aforementioned compromises, our work bridges this gap by experimentally confirming the deniability compromises of the block-layer mobile PDE systems. We have built a mobile device testbed, which consists of a host computing device and a flash storage device. Additionally, we have deployed both the hidden volume PDE and the steganographic file system at the block layer of the testbed and performed disk forensics to assess potential compromises on the raw NAND flash. Our experimental results confirm it is indeed possible for the adversary to compromise the block-layer PDE systems by accessing the raw NAND flash in practice. We also discuss potential issues when performing such attacks in real world. 
\end{abstract}

\section{Introduction}
Mobile computing devices are widely used in our daily life nowadays and, with their increased use, more and more sensitive data are stored and processed in the mobile devices. Therefore, it turns to become an urgent need of protecting those sensitive data, and one of the most critical data security issues is confidentiality. A straightforward approach to protect data confidentiality is to use encryption. Currently, Full Disk Encryption (FDE) has been deployed to the mainstream mobile operating systems including Android~\cite{androidfde} and iOS~\cite{diskencryption}. In FDE, encryption and decryption are completely transparent to users. Without the key, the attacker cannot obtain any knowledge about the original sensitive data. However, FDE cannot defend against a novel coercive attack in which the attacker can force the device owner to disclose the key, and decrypt the ciphertext to obtain the original sensitive data. For example, a journalist or a human rights worker~\cite{yu2014mobihydra,chang2015mobipluto} who is working in a country of conflict or oppression, has captured some sensitive evidence of atrocities and tries to cross the border; to protect the evidence, he/she encrypts the evidence; the border inspector however, may be aware of the ciphertext and force the journalist to disclose the decryption key.

Plausibly Deniable Encryption (PDE) can be used to combat coercive attacks. In PDE, the plaintext is encrypted with a decoy key and a true key. When decrypting the cipher using the decoy key, we will obtain a decoy message and when decrypting the cipher using the true key, we will obtain the true message. Upon being coerced by the attacker, the device owner can only disclose the decoy key and keep the true key secret. In this way, the sensitive data can be protected against the coercive attackers as the attackers cannot notice the existence of the hidden sensitive data. Following the concept of PDE, a large number of PDE systems~\cite{mobiflage_ndss2013, mobiflage_TDSC2014, yu2014mobihydra,chang2015mobipluto,defy_ndss2015,jia2017deftl,chang2018user, chang2018mobiceal,feng2020mobigyges,chen2020poster,chen2020infuse,chen2021usenix,chen2021mobiwear} have been designed for mobile devices. In general, the existing mobile PDE systems can be divided into three categories: C1) block-layer PDE systems; C2) flash translation layer (FTL) PDE systems; and C3) deniability aware flash file systems. 
A majority of the existing mobile PDE systems~\cite{mobiflage_ndss2013, mobiflage_TDSC2014, yu2014mobihydra,chang2015mobipluto,chang2018user, chang2018mobiceal,feng2020mobigyges} belong to the category C1 which deploys PDE on the block layer. The reason is that deploying the PDE on the block layer could be achieved much more easily, resulting in a much better usability. However, the block-layer PDE systems are insecure, because: the hidden sensitive data will leave special traces in the underlying flash memory and such traces cannot be removed by the block-layer PDEs; by having access to the raw flash memory, the adversary may compromise the deniability~\cite{jia2017deftl}. The compromises have been analyzed theoretically by DEFTL~\cite{jia2017deftl}, but none of the existing works have confirmed such compromises experimentally. This work thus aims to bridge this gap by conducting the first experimental study on understanding the deniability compromises of the existing block-layer PDE systems. 

\noindent\textbf{Comparison with DEFTL}. Our work is different from that of the DEFTL~\cite{jia2017deftl} in a few aspects: First, DEFTL theoretically analyzes the potential deniability compromises when deploying the PDE on the block device layer. However, our work experimentally validates the deniability compromises in real-world devices. Especially, we have created a mobile device testbed which includes a host computing device (ARM architecture) and a self-made flash-based block device (using an open-source flash controller and a cheap USB development prototype board). This self-built mobile device follows the architecture of mainstream mobile devices in real world. We then deploy a few representative block-based PDE systems in our testbed, and perform forensic analysis over the raw NAND flash to study the deniability compromises. Second, DEFTL only focuses on the deniability compromises on the PDE systems which use hidden volume technique, but our work assesses both the hidden volume-based and the steganographic file system-based PDE. Third, we have identified extra deniability compromises which have not been discovered in DEFTL.

\section{Background}
\label{sec:background}

\subsection{Flash Memory}
Flash memory especially NAND flash has been used broadly as the external storage of mobile computing devices. Flash memory usually consists of blocks, and each block consists of pages. Compared to traditional hard disk drives (HDD), flash memory exhibits a few unique features: 1) The unit of read and write operations is a page, but the unit of erase operation is a block. 2) A flash page needs to be erased before it can be programmed. 3) Due to the unique features of 1) and 2), the in-place update in flash memory is expensive. Therefore, the flash storage typically performs out-of-place instead of in-place update. 4) Each block in flash memory can only be programmed or erased for a limited number of times. Therefore, program/erase should be distributed evenly across the entire flash.
\subsection{Flash Translation Layer}
To manage flash memory, we can rely on a flash-specific file system like YAFFS\cite{manning2010yaffs} and JFFS~\cite{woodhouse2001jffs}. However, the flash-specific file systems are rarely used today. Instead, a flash translation layer (FTL) has been introduced to handle the special nature of NAND flash hardware, exposing a block access interface externally. Most of the existing mobile devices and flash memory based products (e.g., SD cards, solid state drives, USB drives) use FTL. The core functions implemented by the FTL include garbage collection, wear leveling, and bad block management.

\noindent\textbf{Garbage collection.} During the use of flash memory, a block may contain invalid pages. Garbage collection is often used to reclaim those invalid pages. It usually follows a few steps: 1) choosing a victim block which has the most invalid pages; and 2) coping valid pages in this victim to an empty block; and 3) erasing the victim block. 

\noindent\textbf{Wear leveling.} Each block in flash memory has a limited number of program/erase (P/E) cycles. The main purpose of wear leveling is to distribute P/E cycles evenly across the entire flash. When wear leveling is triggered, flash controller first chooses a block (identified as M) which has the smallest erase count from blocks being used, and then chooses another block (identified as N) which has the largest erase count from the free block pool. Then, data in block M are copied to block N. 

\noindent\textbf{Bad block management.} Blocks may turn bad during the use of flash memory. A bad block table is often maintained to keep track of the bad blocks. If a block turns bad, the data in this block will be copied to an empty block and this bad block will be put to the bad block table. 
\subsection{Plausibly Deniable Encryption}
\begin{figure}[tb]
\centering 
\includegraphics[scale=0.3]{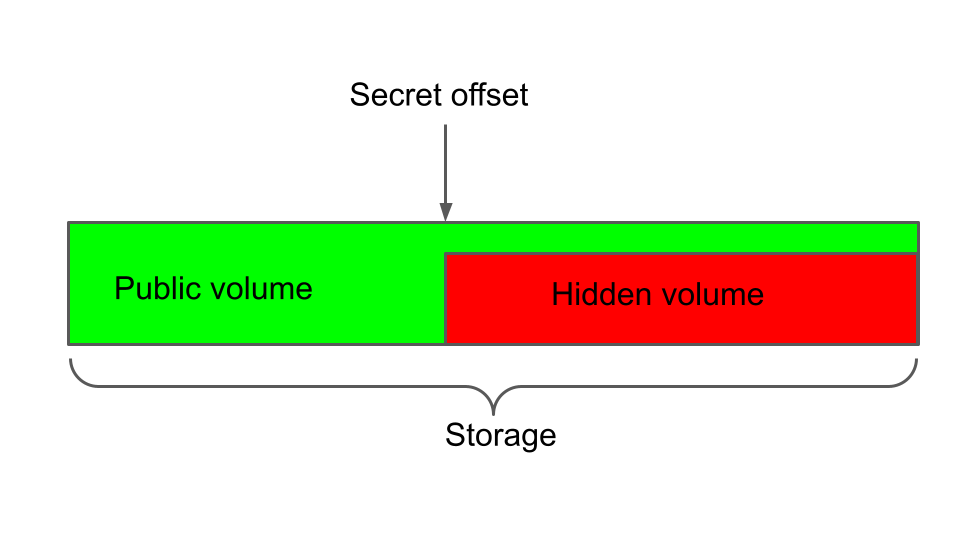}
\vspace{-25pt}
\caption{The hidden volume-based PDE technique.} 
\label{fig:hiddenPDE}
\vspace{-5pt}
\end{figure}

To defend against coercive attacks, plausibly deniable encryption is proposed. Typically, two techniques can be used to implement the PDE system, steganographic file system-based~\cite{stegfs_IH1998} and hidden volume-based technique. Steganographic file system will fill the disk with random data initially, sensitive data are encrypted and stored at a secret address which can be derived from the key. However, steganographic file system suffers from data loss since non-sensitive data may overwrite the sensitive data. To mitigate this problem, several copies of sensitive data are stored in multiple places of the disk to lessen the probability of overwritten. In hidden volume-based technique (Figure~\ref{fig:hiddenPDE}), the entire disk is filled with random data initially. There are two types of volume, public volume and hidden volume. Public volume is encrypted with the decoy key and placed across the entire disk, hidden volume is encrypted with the true key and placed to the end of the disk start from a secret offset. Upon being coerced, the device owner will only disclose the decoy key. The attacker can only decrypt the public non-sensitive data with the decoy key, but he/she cannot notice the existence of hidden volume since he/she cannot differentiate the hidden volume from random data.

\section{Model and Assumptions}
\begin{figure}[h]
\centering 
\includegraphics[scale=0.3]{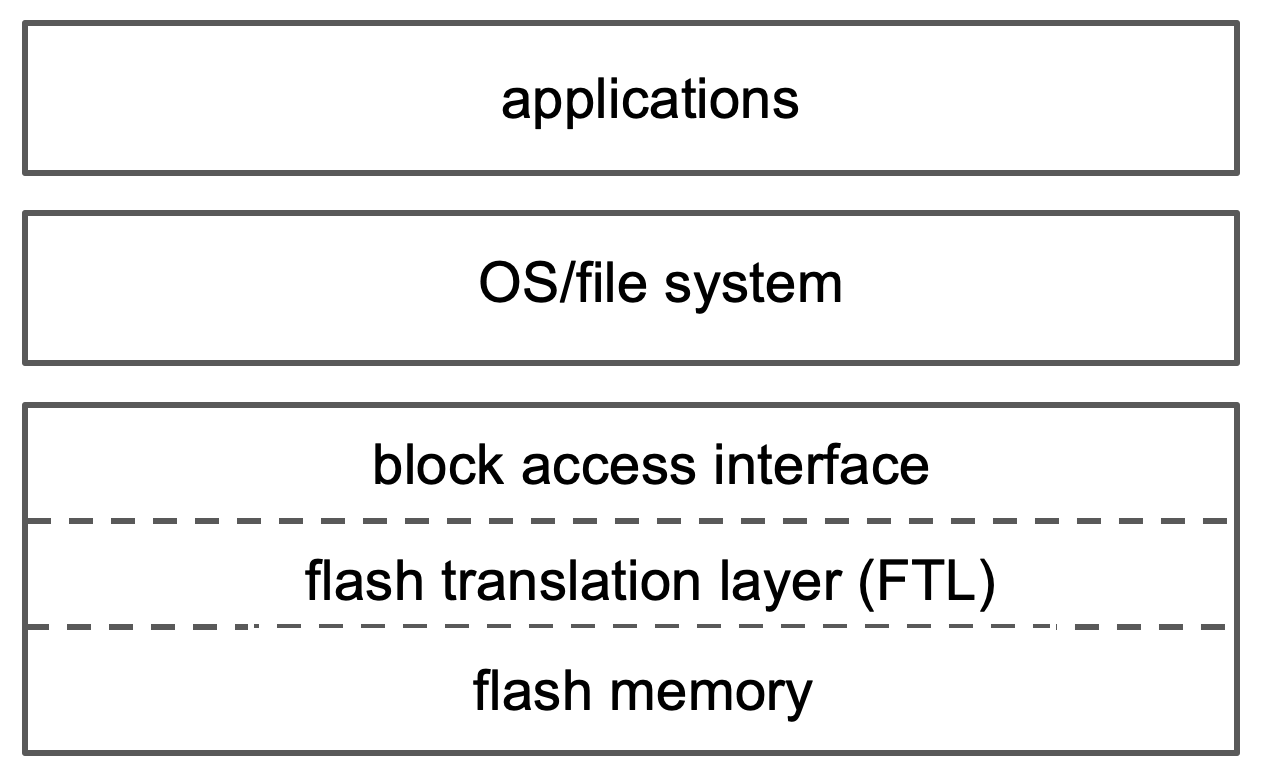}
\vspace{-10pt}
\caption{The storage architecture of main-stream mobile computing devices} 
\label{fig:architecture}
\vspace{-5pt}
\end{figure}

\noindent\textbf{System model}. We consider a mobile computing device which is equipped with flash memory (e.g., UFS cards, eMMC cards, microSD cards, etc) as the external storage. The storage architecture of main-stream mobile devices is shown in Figure~\ref{fig:architecture}. A mobile user directly communicates with apps (e.g., a PDF viewer app) running at the application layer. The OS/file system will manage storage hardware and provide system calls for the applications to access the data stored at the storage hardware. The underlying flash memory storage is typically used in the form of a block device. The FTL will handle special nature of flash memory, exposing a block access interface externally. 

\noindent\textbf{Adversarial model}. We assume the adversary can capture both the victim and his/her mobile device, and coerce the owner to disclose the decryption key. The adversary is rationale and will stop coercing the user once he/she believes that the decryption key is disclosed~\cite{mobiflage_ndss2013,chang2015mobipluto,chang2018mobiceal,jia2017deftl}. Using the disclosed key, the adversary will play with the mobile devices to compromise the PDE. In addition, the adversary can extract the raw image from the flash storage equipped with the victim device and obtain the hardware parameters (e.g., page size and block size) of the underlying flash memory chips. The adversary can then perform forensic analysis on the raw image to identify the existence of PDE.  
\section{Experimentally Attacking The Block-layer PDE Systems}
The hidden volume technique and the steganographic file system are two major techniques which have been leveraged to implement the PDE system at the block layer. We therefore focus on attacking those two types of PDE systems. For each type of PDE systems, we first deploy a representative PDE implementation on a self-built mobile device testbed, and then perform forensic analysis to identify any potential deniability compromises. We mainly concentrate on the deniability compromises in the underlying storage medium, which is typically NAND flash for mobile devices. 

\subsection{Experimental Setup}
\begin{figure}[h]
\centering 
\includegraphics[scale=0.55]{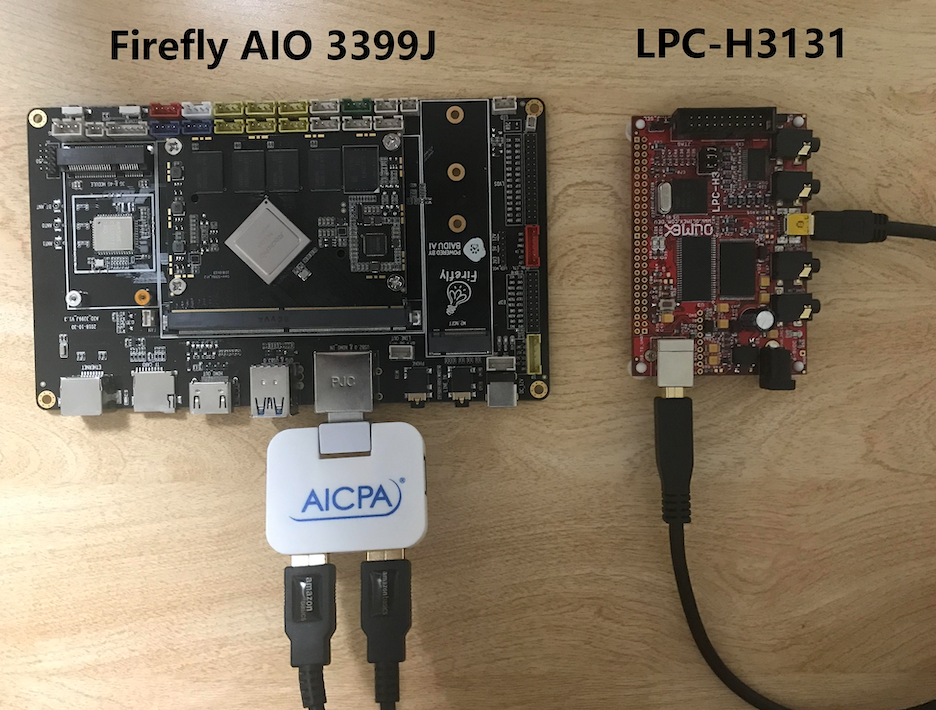}
\caption{A self-made mobile device for our experiment. Firefly AIO-3399J is the host computing device and LPC-H3131 (with flash controller) is the flash-based block device.} 
\label{fig:testbed}
\end{figure}

A challenge faced in our experiment was that, almost every commercially available mobile device (smartphones, tablets, smart watches, or the recent IoT devices like smart home assistants) uses a well encapsulated flash-based block device, e.g., UFS cards, eMMC, microSD cards. 
To facilitate our attacks, we have built a mobile device testbed, which consists of a flash-based block device and a host computing device (Figure~\ref{fig:testbed}). The flash-based block device was built by porting~\cite{tankasala2020step} an open-sourced flash controller OpenNFM~\cite{opennfm} to a USB header development prototype board LPC-H3131~\cite{LPC-H3131} (ARM9 32-bit ARM926EJ-S, 180Mhz, 32MB RAM, and 512MB NAND flash). The host computing device was an embedded development board, Firefly AIO-3399J (Six-Core ARM 64-bit processor, 4GB RAM, and Linux kernel 4.4.194). This mobile device testbed shares a similar architecture with mainstream mobile devices in real world.

We then deployed a block-based PDE system in the host computing device. For the hidden volume-based PDEs, we deployed VeraCrypt~\cite{VeraCrypt}, a fork of the discontinued TrueCrypt project. Note that a large number of PDE systems deployed on the block layer (including PDE systems~\cite{truecrypt,ccs2014} designed for PCs as well as PDE systems~\cite{mobiflage_ndss2013,mobiflage_TDSC2014,yu2014mobihydra,chang2015mobipluto,chang2018user,chang2018mobiceal} designed for mobile devices) have utilized the hidden volume technique, and our attack can be applied to most of them. For the steganographic file systems, we deployed stegfs~\cite{url2}, a recent open-source implementation of steganographic file systems~\cite{stegfs_IH1998,stegfs_IH1999,stegfs_ICDE2003} in user space\footnote{Note that the original implementation of StegFS~\cite{stegfsimplement, stegfs_IH1998,stegfs_IH1999} was done in 1999 for Ext2, and has not been updated since then.}. For each deployed PDE system, we used it, and then extracted the image of the raw NAND flash for further forensic analysis. 

\subsection{Experimental Attacks}
\label{lab: results}
\subsubsection{Experimentally Attacking the Hidden Volume-based PDEs}
We deployed VeraCrypt~\cite{VeraCrypt} in the host computing device, and manually created both a public and a hidden volume via VeraCrypt. The public volume occupies the entire disk (i.e., the flash-based block device built by porting OpenNFM to LPC-H3131) and the hidden volume is 200MB in size. The file system deployed in the public volume was exFAT, which writes data sequentially from the beginning of the disk to avoid overwriting the hidden volume stored stealthily in the second half of the disk. We also deployed exFAT in the hidden volume.  We performed three tests to simulate behaviors of a device owner as follows:

 \noindent\textbf{Test \#1}: We entered the public mode, and wrote non-sensitive data in the public volume. The size of non-sensitive data was less than the size of one block (128 KB) in flash memory (e.g, the size of non-sensitive data can be 2KB). Then, we exited the public mode, entered the hidden mode, and wrote sensitive data to the hidden volume. The size of sensitive data is similar to the non-sensitive data in the public volume (e.g., 2KB). We repeated those operations for several times. This behavior is reasonable. For instance, the user may write a short article (a few KBs in size) to the public volume and store a small secret audio record (a few KBs in size) in the hidden volume. 

 \noindent\textbf{Test \#2}: We entered the public mode and wrote non-sensitive data to the public volume. The size of non-sensitive data should be larger than the size of one block (128 KB). For example, we can write 150 KB non-sensitive data to the public volume. 
Then, we exited the public mode, entered the hidden mode, and wrote sensitive data to the hidden volume. This behavior is reasonable. For instance, the user may store a video in the public volume (a few MBs in size) while storing a small secret audio record (a few KBs in size) in the hidden volume. 

\noindent\textbf{Test \#3}: We entered the hidden mode, and wrote sensitive data to the hidden volume. The size of the sensitive data should be larger than the size of two blocks (256 KB). 
Then, we modified the data in a few randomly chosen locations. This behavior is reasonable. For instance, the user may store a secret document (a few MBs in size) in the hidden volume and modify the document later.

After each test, we analyzed the corresponding flash memory image. Note that the coercive adversary should have access to the decoy key.

\begin{figure}[tb]
\centering 
\includegraphics[scale=0.37]{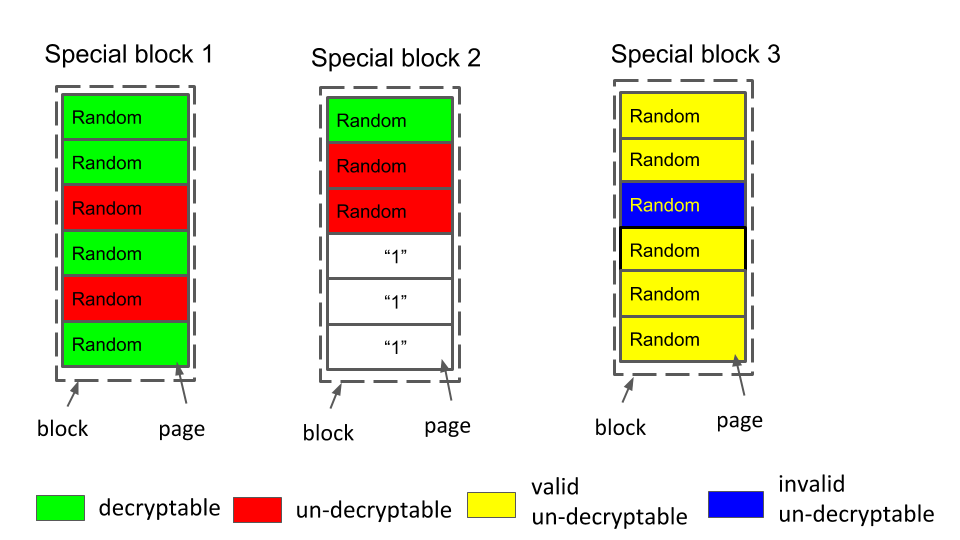}
\vspace{-5pt}
\caption{Special blocks observed in raw NAND flash.} 
\label{fig:specialblock}
\end{figure}

From the image obtained after test \#1, we detected one type of special blocks (i.e., ``special block 1'' of Figure~\ref{fig:specialblock})
which is filled with random data and, a portion of pages among this block cannot be decrypted successfully. \textit{Without the PDE deployed}, there are only two possibilities for a block filling with random data: 1) All data stored in it can be decrypted successfully, i.e., the block is filled with public data. 2) All data stored in it cannot be decrypted successfully, i.e., the block is completely occupied by random data filled initially. However, \textit{with the PDE deployed}, a few pages of the block may be occupied by the hidden data and cannot be decrypted. Therefore, the existence of ``special block 1'' indicates the device owner has entered the hidden mode and committed hidden sensitive data to the external storage before.

From the image obtained after test \#2, we detected the second type of special blocks (``special block 2'' of Figure~\ref{fig:specialblock}) which has a few pages in the beginning storing random data and the remaining pages filling with all `1' bits; some of the random data cannot be decrypted. \textit{Without the PDE deployed}, a block is erased and then partially used by the public data which are all decryptable. However, \textit{with the PDE deployed}, some pages in this block may be used by the hidden data and hence cannot be decrypted. Therefore, the existence of ``special block 2'' also indicates the device owner has committed hidden sensitive data to the external storage before.

From the image obtained after test \#3, we detected the third type of special blocks (``special block 3'' of Figure~\ref{fig:specialblock}) is completely filled with undecryptable random data, but some of them (i.e., in arbitrary locations across the block) are marked as invalid. \textit{Without the PDE deployed}, for the block occupied by randomness filled initially, there are two possibilities: 1) All the pages in this block have been invalidated. 2) The pages at the beginning of the block have been invalidated. Due to the use of exFAT, the system will not write an arbitrary location among the disk, and correspondingly, the FTL will not invalidate an arbitrary page in this block. However, \textit{with the PDE deployed}, this block may be also used by the hidden data, and arbitrary pages across the block may have been invalided by the hidden mode. Therefore, the existence of ``special block 3'' indicates the existence of the hidden mode.

\subsubsection{Experimentally Attacking The Steganographic File System-based PDEs} 
We deployed stegfs~\cite{url2} in the host computing device. Note that the steganographic file system works differently from the hidden volume technique that: the file system is initially filled with randomness and the sensitive data are encrypted with secret keys and stored at random locations of the entire disk; to prevent the hidden data from being overwritten by the public data, it usually needs to maintain a few copies of the hidden data among the entire disk. We performed one test to simulate the behavior of a device owner as follows:
 We first mounted the FAT file system on the flash device, and wrote a certain amount of public non-sensitive data. Then, we manually mounted the steganographic file system, and wrote a few sensitive data. This behavior is reasonable. For instance, the user may first store a few non-sensitive images to the disk via the public file system and then store some secret passwords to the disk via steganographic file system.

\begin{figure}[tb]
\centering 
\includegraphics[scale=0.35]{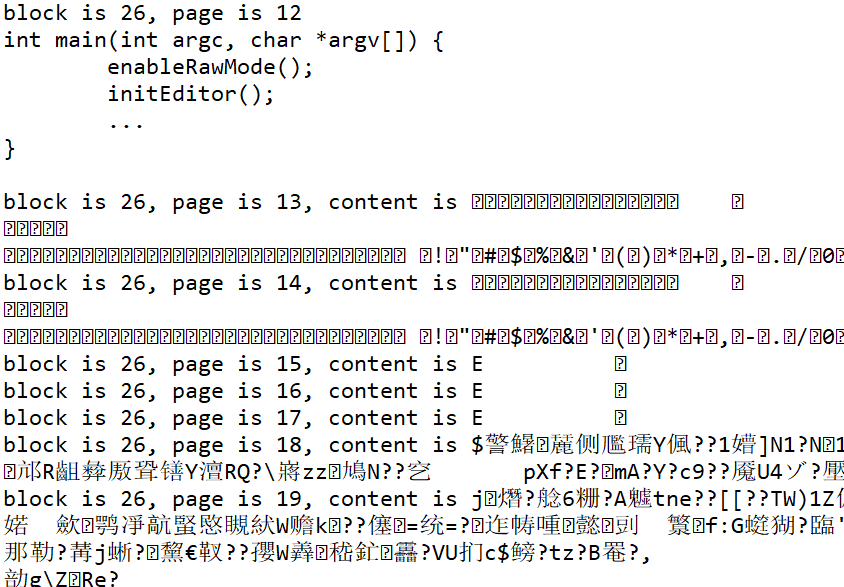}
\caption{A snapshot from a flash block in the attack 2} 
\label{fig:attack-stegfs}
\end{figure}

After the test, we extracted the corresponding flash memory image and analyzed it. We detected a few special traces due to the existence of the PDE: 1) Public data and random data are interleaved across the entire flash. However, without existence of hidden sensitive data, the distribution of the data across the flash should be public data followed by random data. This is because, the steganographic file system fills random data across the entire flash initially and, since the FTL uses log-structured writing, regardless how the file system writes public data, it will always program flash blocks from the beginning. 2) Public data and random data share the same flash block (Figure~\ref{fig:attack-stegfs} shows a snapshot we obtained from one flash block). 
However, \textit{without the existence of the hidden sensitive data}, the distribution of data in a flash block should be either i) public data, followed by all `1' bits, or ii) all public data. This is because, without existence of hidden data, each time when the FTL writes public data but cannot find empty pages, it will erase a flash block, and write public data sequentially from the beginning of the block due to the use of log-structured writing; if any pages in this block has not been filled, they remain empty and contain all `1's. Therefore, the existence of the aforementioned special traces indicates the existence of the hidden sensitive data, leading to the deniability compromise.

\section{Discussion}
\noindent\textbf{Assessing the difficulty of performing our attacks}. 

To compromise the deniability by having access to the raw flash memory, the adversary needs to tackle two issues: 1) how to extract an image from the NAND flash memory given a victim mobile device, and 2) how to perform forensic analysis over the raw flash memory data. For the first issue, Breeuwsma et al.~\cite{breeuwsma2007forensic} introduced a few low-level data acquisition methods for flash memory, including flasher tools, using an access port commonly used for testing and debugging, etc. Chen et al.~\cite{chen2021usenix} mentioned a method of obtaining raw data from SSDs ``by opening the covers and directly reading the memory chips with cheap off the shelf readers''. For the second issue, the adversary can use the existing digital forensic tools available on the market (e.g., Photorec~\cite{photorec}, TestDisk~\cite{testdisk}, Foremost~\cite{foremost}, Cellebrite UFED~\cite{cellebrite}/ Inspector~\cite{cellebriteinspector}, etc) or develop new special tools to analyze the captured image.

\noindent\textbf{Implications of our experimental attacks}. Our attacks performed in this work confirm that it is indeed feasible to compromise the block-based PDE systems in practice. Our results further justify that the deniability compromise in the lower storage medium is indeed a significant issue and should be considered seriously when designing any future PDE systems for mobile computing devices. An immediate remediation would be moving the entire PDE system design to the flash translation layer (FTL)~\cite{jia2016nfps,chen2021usenix} which however, would not be a good solution as it will impose a large burden on the FTL firmware. In addition, as the PDE integrated in the FTL firmware is far away from the user applications, making it user unfriendly. It is still unclear how to design a PDE system which is 1) secure (i.e., eliminating deniability compromises in the flash memory), and 2) keeping the FTL lightweight, and 3) user-friendly which is still an open problem in the literature. 
 
\noindent\textbf{Other attacks on the PDE systems}. 
This work only focuses on the single-snapshot attack in which the adversary is only allowed to have access to the victim device once. A stronger adversary may conduct the multiple-snapshot attack by periodically accessing to the victim device~\cite{chen2020towards,chen2020poster,chen2021usenix}. By capturing different snapshots of the external storage over time and comparing the different snapshots, the adversary will detect changes of the hidden sensitive data, compromising the deniability. For example, if the hidden volume technique is used, by comparing different snapshots, the adversary may observe data changes performed in the space which is claimed empty but actually hides the hidden volume. Some of mitigation strategies can be accompanying public writes with dummy writes and hiding the sensitive data into the dummy writes~\cite{chen2020towards,chen2020poster}, or using the WOM (write-once memory) code to encode the hidden data in a public cover~\cite{chen2021usenix}. 
In addition, this work only focuses on the deniability compromises in the external storage, but hidden sensitive data may leave traces in the internal memory, and such traces may be extracted by the adversary by performing memory forensics~\cite{burdach2006physical}. One potential solution is to power-off the device each time after quitting the hidden mode in which the user can manage the hidden sensitive data. Another solution could be leveraging trusted execution environments (TEE) so that the memory used to process the hidden sensitive data can be protected, avoiding being accessed by the adversary. 

\section{Related Work}
Plausibly deniable encryption is used to defend against coercive attacks. Currently, there are two major techniques which can be leveraged to implement a PDE system at the block layer, namely, the hidden volume technique and the steganographic file system.

\noindent\textbf{The hidden volume-based PDE systems}. Skillen et al. proposed Mobiflage~\cite{mobiflage_ndss2013, mobiflage_TDSC2014}, the first PDE system designed for mobile devices. Mobiflage has two versions: the first version assumes the existence of an FAT32 SD card, and the second version releases the aforementioned assumption by using a modified Ext4 file system. Yu et al. proposed~\cite{yu2014mobihydra} MobiHydra, which allows the user to switch from the public to the hidden mode without rebooting the device. MobiHydra also mitigates the booting-time attack faced by Mobiflage. Chang et al. designed Mobipluto~\cite{chang2015mobipluto}, a file system friendly PDE system which allows any block-based file systems to be used in the public volume, by smartly integrating the hidden volume technique with thin provisioning. Chang et al. further extended the hidden volume technique to combat the multi-snapshot adversary by introducing dummy writes on the block layer~\cite{chang2018mobiceal}. 

\noindent\textbf{The steganographic file systems}. 
Anderson et al.~\cite{stegfs_IH1998} proposed the first steganographic file system to achieve PDE. The secret data are hidden among the randomness. The system maintains several copies of secret data to mitigate overwrite problem. Inspired by~\cite{stegfs_IH1998}, McDonald et al.~\cite{mcdonald1999stegfs} designed a more practical and efficient steganographic file system. Instead of leveraging a separate partition of a harddisk, secret files are hidden in the unused blocks of a partition that also contains normal files. Pang et al.~\cite{pang2003stegfs} then proposed StegFS. The blocks that store encrypted secret data are distributed across the entire storage. The locations of those blocks cannot be obtained without the keys. In this way, the attacker cannot notice the existence of hidden data. Peters et al.~\cite{defy_ndss2015} proposed DEFY, the first deniable file system designed for log-structured storage. Barker et al.~\cite{barker2020artifice} proposed Artifice, which addresses problems regarding flash storage devices and multiple snapshot attacks through
comparatively simple block allocation schemes and operational security.
\section{Conclusion}
In this work, we have experimentally confirmed the deniability compromises of the block-layer PDE systems deployed on the mobile computing devices. Our work conducts the first experimental attacks by 1) deploying both the hidden volume-based PDE and the steganographic file system on the block layer of a mobile device testbed, and 2) allowing the adversary to have access to the flash memory and to perform forensic analysis over the raw flash memory data. Our results strengthen the necessity of taking care of the deniability compromises in the lower storage layer when designing any future PDE systems for mobile devices. 

\bibliographystyle{plain}
\bibliography{reference}
\end{document}